\documentclass[12pt,letterpaper,twoside]{article}

\usepackage[margin = 1in]{geometry}

\usepackage{amsmath, amssymb, amsfonts, amsthm}
\usepackage{bm}
\usepackage{mathtools}
\usepackage{breqn}
\usepackage{mathrsfs}
\usepackage{bbm}
\usepackage{dsfont}
\usepackage{derivative}

\usepackage[dvipsnames]{xcolor}
\usepackage{graphicx}
\usepackage[export]{adjustbox}
\usepackage{caption}
\usepackage{subcaption}
\usepackage{float}
\usepackage{rotating}
\usepackage{array}
\usepackage{booktabs}
\usepackage{multirow}
\usepackage{makecell}

\usepackage{tikz}
\usetikzlibrary{automata, fit, shapes.geometric}

\usepackage{natbib}
\usepackage{authblk}

\usepackage[title]{appendix}

\usepackage{xr}

\usepackage{fancyhdr}
\makeatletter
\newcommand*{\addFileDependency}[1]{%
  \typeout{(#1)}%
  \@addtofilelist{#1}%
  \IfFileExists{#1}{}{\typeout{No file #1.}}%
}
\makeatother

\usepackage[colorlinks=true, linktocpage=true]{hyperref}
\hypersetup{
  colorlinks=true,
  linkcolor=Blue,
  citecolor=BrickRed,
  urlcolor=Blue,
  filecolor=Blue
}

\usepackage[capitalise]{cleveref}

\usepackage[colorinlistoftodos,prependcaption]{todonotes}

\allowdisplaybreaks

\emergencystretch=\maxdimen
\newcommand{\spacingset}[1]{%
  \renewcommand{\baselinestretch}{#1}\small\normalsize
}

\begin{document}

\spacingset{1}


\title{\bf JASPER: Joint Bayesian Analysis of Spatial Expression via Regression}
\date{}

\author[1, *]{Pritam Dey}
\author[1]{Rajarshi Guhaniyogi}
\author[1]{Yang Ni}
\author[1]{Bani K. Mallick}

\affil[1]{\footnotesize Department of Statistics, Texas A\&M University, College Station, TX}
\affil[*]{\footnotesize All correspondence should be addressed to pritam.dey@tamu.edu}
\maketitle

\bigskip
\begin{abstract}
Spatially resolved transcriptomics is a fast-developing set of technologies that enables the measurement of localized gene expression across spatial locations in a sample. Detecting spatially varying genes is critical for analyzing such data, yet existing methods often fail to account for inter-gene correlations, leading to inflated false positive and false negative rates. Additionally, most prominent methods rely on predefined spatial covariance kernels, making them sensitive to the complexity of spatial expression patterns. Motivated by a human breast cancer dataset, we address these limitations in existing literature through JASPER (Joint Bayesian Analysis of SPatial Expression via Regression), a Bayesian framework that jointly models spatial expression patterns across multiple genes using a spatial basis function regression approach. We demonstrate the superior performance of JASPER compared to existing methods in several real-world spatial transcriptomic datasets and supporting simulation experiments. JASPER identifies genes with stronger spatial correlation and greater biological relevance, as validated by overlap comparison, enrichment analysis, and pathway analysis using independent biological databases. Our results highlight the ability of JASPER to improve the statistical and biological interpretability of spatial transcriptomics data, making it a powerful tool for uncovering spatial gene expression patterns in complex biological systems. Details on the exact posterior inference algorithm, simulation settings, and additional results are provided in the Supplementary Materials.
\end{abstract}

\sloppy \noindent%

\vfill

\newpage
\spacingset{1.8} 

%


\maketitle


%

\section{Introduction}
\label{section:introduction}

Spatially resolved transcriptomics (SRT) is a set of rapidly advancing technologies that measure mRNA gene expression within the spatial context of tissue architecture~\citep{srt2, srt3}. Unlike traditional transcriptomics methods that only measure average gene expression for bulk samples, SRT allows researchers to investigate where specific genes are expressed within a tissue sample. This spatial resolution is critical for understanding cellular architecture, tissue organization, and disease microenvironments. Broadly speaking, SRT technologies can be categorized as imaging-based and next-generation sequencing (NGS) based approaches. Imaging-based methods, such as Multiplexed Error-Robust Fluorescence In Situ Hybridization (MERFISH)~\citep{merfish1} and Sequential Fluorescence In Situ Hybridization (SeqFISH)~\citep{seqfish} utilize spatially resolved imaging to quantify mRNA molecules within intact tissues at single-cell resolution. In contrast, NGS-based approaches, including Spatial Transcriptomics (ST)~\citep{spatialtranscriptomics}, 10x Genomics Visium, Slide-seq~\citep{slideseq} and Spatially Resolved Transcript Amplicon Readout Mapping (STARmap)~\citep{starmap} rely on spatially barcoded oligonucleotides to capture and sequence transcripts in spatially defined regions of the tissue. Comprehensive reviews of various SRT platforms can be found in~\cite{srt-methods-1, srt-methods-2}. The emergence and popularity of these diverse platforms have necessitated the development of newer statistical methodologies tailored to model the highly (spatially) structured biological datasets obtained from SRT studies.

With SRT data, we can identify genes that have spatially dependent expression profiles. These genes, commonly referred to as spatially varying genes (SVGs), can provide researchers with a host of valuable biological insights, including knowledge of tissue heterogeneity, characterization of underlying mechanisms and pathways, discovery of clinical biomarkers associated with specific tissue regions or cell types, and identification of potential therapeutic targets for personalized treatments. These critical scientific potentials underscore the importance of identifying SVGs as a key first step in analyzing SRT datasets. Consequently, there is an acute necessity for fast, accurate, and robust statistical methodologies to address the SVG detection problem.

\subsection{Related Work}
\label{sub:related-work}

The SVG detection problem shares similarities with the problem of identifying differentially expressed (DE) genes for microarray data \citep{geneselectionmicroarrayrf}. Both problems aim to identify genes that exhibit certain variation in expression. However the key difference is that while DE gene analyses aim to find genes whose bulk expression (i.e., expression over a large number of cells) vary across discrete sample groups such as treatments, SVG detection aims to identify genes that exhibit variation in expression according to spatial location in a tissue or cell-culture at a fine resolution (typically for single cells or a small number of neighboring cells).

A variety of statistical and computational models have been developed to address the SVG detection problem. These methods are based on a variety of frameworks including (a) spatial metrics like Moran's I statistic \citep{moranI} or Geary's C \citep{gearyC} for ranking genes according to spatial correlations, (b) model-based hypothesis testing frameworks such as SpatialDE \citep{spatialde}, SPARK \citep{spark}, SOMDE \citep{somde}, nnSVG \citep{nnsvg}, and Trendsceek \citep{trendsceek}, and (c) model-free approaches such as SPARK-X \citep{sparkx} and sepal \citep{sepal}. A comprehensive review of various SVG selection methods can be found in \cite{svgdetectreview}. However these methods all approach the SVG detection problem from a screening perspective, i.e., they consider genes one-by-one and test if they exhibit spatially varying expression patterns. While this gene-by-gene screening approach is highly scalable, these approaches ignore the co-expression of genes, i.e., the fact that the expression of a gene may be regulated by the expression of other genes or underlying biological states. To the best of our knowledge, no existing SVG detection method take these effects into consideration. We address this gap by approaching the problem from a variable selection perspective, i.e., we construct a model of correlated expressions for multiple genes at multiple spatial locations and use it to identify those genes whose expression is spatially varying. This perspective allows us to borrow tools from the rich variable selection and spatial variable selection literature, which we review below.

Variable selection enjoys a central position in the statistics and machine learning literature. Traditional variable selection methods consist of both frequentist approaches such as subset selection \citep{miller2002subset}, penalized regression \citep{tibshirani1996lasso, hoerl1970ridge} and model selection criterion such as AIC/BIC \citep{akaike1974aic, schwarz1978bic} and Bayesian approaches such as spike-and-slab priors \citep{george1993gibbs} and shrinkage priors \citep{park2008bayesianlasso, carvalho2010horseshoe}. Variable selection in classification and regression models improves interpretability, prediction accuracy, and model robustness. While these methods select subsets of variables that best explain a response variable globally, specialized methods exist for spatial variable selection, which aim to identify variables that best explain the spatial variability of a given variable. 

Driven by recent advances in environmental science, neuroimaging, and genetics, spatial variable selection methods have become immensely popular in modern spatial statistics literature. A typical albeit computationally intensive approach is recursive feature elimination with spatial cross-validation \citep{meyer2019importance, ketu2022spatial}. Alternative strategies include geographically weighted regression techniques and their variants \citep{brunsdon1998geographically, wheeler2021geographically} and spatially varying coefficient models \citep{gelfand2003spatial, guhaniyogi2022distributed}, which address spatially varying associations of predictors and responses. Geographically weighted regression approaches have been extended by allowing for site-specific \citep{lum2012bayesian, scheel2013bayesian} inclusion of predictors and joint site-specific predictor selection and regression coefficient estimation \citep{boehm2015spatial,choi2018bayesian}. In neuroimaging studies, spatial variable selection priors are often utilized to identify brain regions activated by stimuli \citep{lee2014spatial, zhang2014spatio} and to estimate relationships between image voxels and phenotypes. Some notable approaches to discern the relationship between image voxels and phenotypes employ basis function representations to model spatially varying image coefficients \citep{li2015spatial}, as well as place thresholded Gaussian process priors \citep{zeng2024bayesian} on spatially-varying image coefficients. \cite{nakajima2013bayesian} proposed a sparsity-inducing hard threshold prior for longitudinal data, providing room to extend such framework for similar spatial variable selection methods. Lastly, there is related literature on functional variable selection in function-on-scalar, scalar-on-function, and function-on-function regression frameworks \citep{chen2016variable, reiss2017methods}, which consider images as multi-dimensional functions. These frameworks popularly employ basis functions and functional principal components analysis with regularization on their coefficients, as detailed in \cite{aneiros2022variable} and references therein. 

\subsection{Our contributions}
\label{sub:contributions}

In this article, we introduce JASPER (Joint Bayesian Analysis for SPatial Expression via Regression), a fully Bayesian method for jointly detecting SVGs in SRT data. JASPER incorporates the following key features:
%
\begin{itemize}
    \item \textbf{Joint modeling of genes:} JASPER simultaneously models multiple genes, capturing gene co-expression structures alongside spatial expression patterns. This contrasts with existing methods that analyze genes individually, potentially overlooking important spatial dependencies.
    
    \item \textbf{Spatial basis function regression:} Instead of relying on pre-specified spatial covariance kernels, JASPER employs a spatial basis function regression framework, allowing for flexible modeling of diverse spatial patterns without the constraints of Gaussian process-based approaches.
    
    \item \textbf{Modeling of expression counts:} JASPER models gene expression counts using a negative-binomial distribution, avoiding the need for pre-normalization. Compared to Poisson-based models, the negative binomial formulation better accounts for overdispersion and zero inflation, both common in SRT data. 
\end{itemize}
%
%
Although JASPER is adaptable to data from diverse SRT platforms as we demonstrate in this article, we were motivated by a recently published human breast cancer dataset~\citep{iimpact} obtained using the 10X Visium platform. In Section \ref{section:data-description}, we highlight the key features of JASPER within the context of this dataset, with a special focus on its complex tissue structures and the spatial heterogeneity of gene expression patterns. We illustrate how these challenges reduce the effectiveness of existing methods and showcase how our methodological innovations make JASPER a more robust and biologically meaningful alternative compared to the state-of-the-art competing approaches. Extensive simulations in Section \ref{section:simulations} further validate our claims, showing that JASPER consistently identifies SVGs with greater precision than competing methods.

Section \ref{section:methods} formally introduces the JASPER model and outlines a partially collapsed, data-augmented Gibbs sampler for Bayesian inference. Section \ref{section:simulations} validates JASPER through simulation studies. In Section \ref{section:real-data}, we present the results of applying our method alongside two state-of-the-art alternatives on the human breast cancer dataset, highlighting key biologically significant insights uniquely recovered by JASPER. To demonstrate the versatility of our approach, we apply it to two additional datasets obtained from different SRT platforms than the motivating breast cancer dataset. Finally, we conclude with a discussion of the key insights provided by our model and propose several potential directions for future research in Section \ref{section:discussion}.


\section{Human Breast Cancer Dataset}
\label{section:data-description}

\begin{figure}[H]
    \centering
    \includegraphics[width=\textwidth]{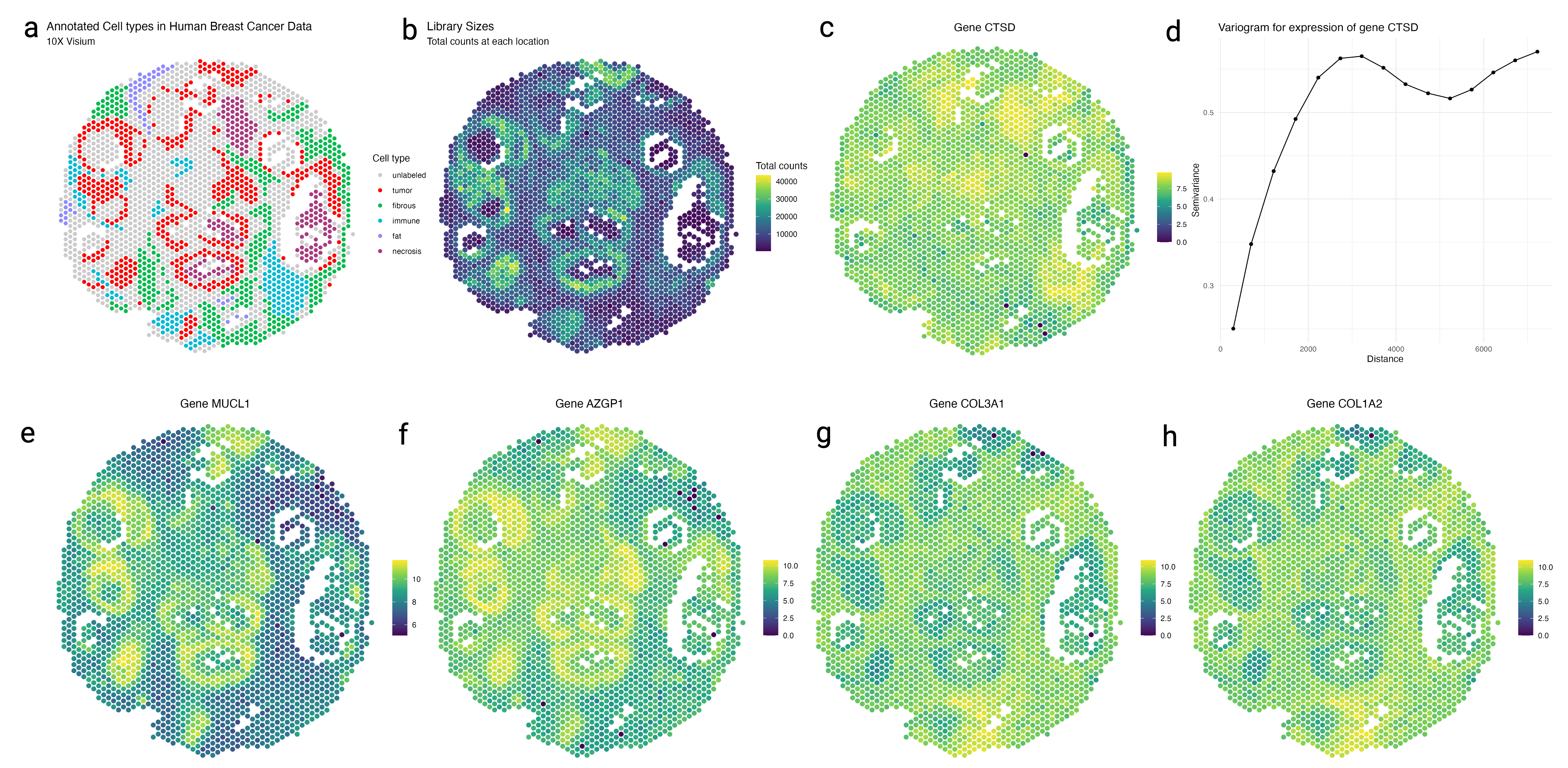}
    \captionsetup{font={stretch=1.1}}
    \caption{Human Breast Cancer data obtained using the 10x Visium pipeline. (a) Expert Annotatations of Spots, (b) Library Sizes, (c-d) Normalized gene expression and variogram for gene CTSD with empirical evidence of spatial non-stationarity, (e-h) Two pairs of highly correlated genes (correlation for left pair: 0.926, correlation for right pair: 0.867)}
    \label{fig:bc-descriptive}
\end{figure}

The human breast cancer dataset, introduced in~\cite{iimpact}, comprises of mRNA gene expression counts for 17,651 genes measured at 2,518 spatially indexed locations within a section of human breast tissue diagnosed with invasive ductal carcinoma. The dataset was generated using the 10x Genomics Visium platform, a widely adopted commercial NGS technology. In addition to spatially resolved gene expression measurements, the dataset includes expert annotations from pathologists for a subset of spatial locations, offering additional histo-pathological insight. As shown in Panel (a) of Figure \ref{fig:bc-descriptive}, these annotations vividly highlight the spatial heterogeneity inherent in the tissue structure.

An important characteristic of SRT datasets in general and this breast cancer dataset in particular is that the expressions of several genes are highly correlated with each other (Figure \ref{fig:bc-descriptive}, panels (e-h)). Existing SVG detection methods fail to account for these correlative structures owing to their gene-by-gene screening perspective. In contrast, JASPER, models these correlations explicitly using a joint Bayesian model, thus approaching the SVG detection problem though the lens of variable selection. This joint modeling approach allows JASPER to share expression information among co-expressed genes and enables precise identification of SVGs by isolating spatially dependent variations from other sources of gene expression variability. Through simulation experiments (in section \ref{sub:simulations-coexp}), we demonstrate how JASPER is able to achieve much lower false positive and false negative rates compared to existing methods. Further these simulations indicate that this improvement in SVG detection accuracy is more pronounced when genes are more strongly correlated. 

As demonstrated by the expert manual annotations in Panel (a) of Figure \ref{fig:bc-descriptive}, the complex architecture of the tissue underlying this dataset features a diverse array of cells arranged within non-convex and non-contiguous spatial domains. This structural complexity results in intricate patterns of high and low gene expression, as shown in Panel (e-h) of Figure \ref{fig:bc-descriptive}. Most current model-based SVG detection approaches rely on Gaussian processes to model spatial dependencies. However, these methods are often constrained by their choice of covariance kernels. For instance, commonly used squared-exponential and, more generally, Mat\'{e}rn kernels assume stationarity in Gaussian processes—an assumption frequently violated in SVG expression profiles. Notably, in our breast cancer dataset, visual inspection of the omnidirectional semivariograms reveals clear evidence of spatial non-stationarity -- semi-variance does not stabilize with increasing distance; see one such example in Figure \ref{fig:bc-descriptive}, Panels (c-d). Kernel mis-specifications can significantly reduce the statistical power of SVG detection methods. Some existing methods such as SPARK attempt to circumvent this issue by testing against several different kernel functions. However, their reliance on subjective kernel selection remains a critical limitation. To illustrate the sensitivity to covariance kernels, we conducted a simulation study (Section \ref{sub:simulations-kernel}) where data was generated using a periodic exponential-squared-sine kernel. Under this simulation setting, kernel-based methods like SpatialDE~\citep{spatialde} and SOMDE~\citep{somde}, which utilize squared-exponential kernels, exhibited low true positive rates of 43.2\% and 63.5\%, respectively. In contrast, JASPER avoids the pitfalls of kernel selection by leveraging linear combinations of spatial basis functions. In the same simulation setting, JASPER achieved a true positive rate of 93.5\%, demonstrating its ability to adapt to diverse spatial patterns without explicitly specifying a particular kernel function.

Another important characteristic of this dataset is the pronounced spatial pattern in library sizes, defined as the total observed mRNA counts at each spatial location (Figure \ref{fig:bc-descriptive}, panel (b)). Consequently, SVG detection methods that apply standard library size-based normalization techniques, such as SpatialDE~\citep{spatialde} and SOMDE~\citep{somde}, may inadvertently introduce spurious spatial effects, leading to inflated false positive rates. A common strategy to mitigate these artifacts involves normalization techniques that account for differential gene expression, such as trimmed mean of M-values (TMM)~\citep{tmm-normalization} and upper-quantile normalization~\citep{upper-quartile}. In contrast, JASPER directly models the raw expression counts using negative binomial distributions, which eliminates the need for pre-normalization while preserving the inherent statistical properties of the gene expression data. This approach also helps avoid potential distortions in the natural relationship between mean expression levels and variance in raw counts, a known issue with traditional normalization techniques~\citep{dontnormalize}.
%
In addition to the breast cancer dataset, we demonstrate the versatility of our proposed methodology by applying it to two additional SRT datasets generated using other popular platforms. The first of these additional datasets was derived from the mouse visual cortex using the STARmap technique~\citep{starmap} and includes expression counts for 160 genes measured at 891 spatially indexed locations. The second is a mouse olfactory bulb dataset obtained using the Spatial Transcriptomics (ST) technology~\citep{spatialtranscriptomics}, and contains expression counts of 16,573 genes measured at 265 spots. In both datasets, JASPER was able to detect more biologically and statistically significant SVGs compared to existing methods, which showcases the versatility of our approach across different SRT technologies.


\section{Methodology}
\label{section:methods}

Consider a typical SRT dataset consisting of counts mapped to $p$ genes labeled $1,2,\dots,p$ observed at $n$ spatial locations (cells/spots) labeled $\bm{s}_1, \bm{s}_2, \dots, \bm{s}_n$ in a 2 dimensional spatial domain (the biological sample, such as a slide). Let $C_j(\bm{s}_i)$ denote the count for gene $j$ at location $\bm{s}_i$ and let $\bm{C}(\bm{s}_i) = \left(C_1(\bm{s}_i), C_2(\bm{s}_i), \dots, C_p(\bm{s}_i)\right)^{\top}$ denote the vector of gene expression counts for all genes at location $\bm{s}_i$. Throughout this article, $\bm{I}_k$ denotes the $k \times k$ identity matrix, $\bm{1}_k$ and $\bm{0}_k$ denote the $k$-dimensional vectors containing all 1's and 0's, respectively, and $\delta_{\bm{v}}$ is a point mass at $\bm{v}$. 

\subsection{Negative Binomial Model for Observed Counts}
\label{sub:neg-binom-methods}

We model the observed gene expression count $C_j(\bm{s}_i)$ using a negative binomial distribution conditional on the latent continuous gene expression surface $Y_j(\bm{s}_i)$:
\begin{equation}
\label{eq:negative-binomial-model}
\begin{aligned}
f\left(C_j(\bm{s}_i)\mid Y_j(\bm{s}_i)\right)
&=
\frac{\Gamma\{C_j(\bm{s}_i)+\phi_j\}}
{\Gamma(\phi_j)C_j(\bm{s}_i)!}
\left(
\frac{\mu_j(\bm{s}_i)}
{\mu_j(\bm{s}_i)+\phi_j}
\right)^{C_j(\bm{s}_i)}
\left(
\frac{\phi_j}
{\mu_j(\bm{s}_i)+\phi_j}
\right)^{\phi_j},\\
\mu_j(\bm{s}_i)
&=
N_i\exp\{Y_j(\bm{s}_i)\},
\end{aligned}
\end{equation}
where $\phi_j>0$ is a gene-specific dispersion parameter of the negative-binomial count model and $N_i$ adjusts for differences in sequencing depth across locations. Typical choices for $N_i$ include the observed library sizes,
$N_i=\sum_{j=1}^p C_j(\bm{s}_i)$, or normalized library sizes obtained from procedures such as TMM~\citep{tmm-normalization}. In practice, we recommend using normalized library sizes, particularly when sequencing depth exhibits spatial variation. Under \eqref{eq:negative-binomial-model}, $\mathbb{E}[C_j(\bm{s}_i)\mid Y_j(\bm{s}_i),\phi_j] = \mu_j(\bm{s}_i)$ and $\mathrm{Var}\{C_j(\bm{s}_i)\mid Y_j(\bm{s}_i),\phi_j\} = \mu_j(\bm{s}_i) + \tfrac{\mu_j^2(\bm{s}_i)}{\phi_j}$, which naturally accommodates overdispersion in the observed counts.

\subsection{Low-Rank Spatial Factor Regression Model for Latent Expression}
\label{sub:low-rank-spatial-model}

Consider a set of $K$ spatial basis functions defined over the spatial domain of the locations $\bm{s}_1, \bm{s}_2, \dots, \bm{s}_n$ denoted as $X_k(\cdot)$ for $1 \leq k \leq K$. We model the latent expression level $Y_j(\bm{s}_i)$ corresponding to gene $j$ at location $\bm{s}_i$ as: 
\begin{equation}
\label{latent-low-rank-model}
Y_j(\bm{s}_i) =
\alpha_j
+ \bm X(\bm{s}_i)^\top \bm\beta_j
+ \bm f_i^\top \bm\lambda_j
+ \epsilon_{ij},
\qquad
\epsilon_{ij} \sim \mathcal N(0,\psi_j),
\end{equation}
where $\bm X(\bm{s}_i)=\left(X_1(\bm{s}_i),\dots,X_K(\bm{s}_i)\right)^\top$ denotes the vector of basis functions evaluated at location $\bm s_i$. Further, $\alpha_j$ and $\bm\beta_j\in\mathbb R^K$ denote the intercept and spatial basis coefficient vector for gene $j$, respectively, while $\bm f_i,\bm\lambda_j\in\mathbb R^r$ represent the location-specific latent factor score and the corresponding gene-specific factor loading vector, respectively. 
The low-rank term $\bm f_i^\top \bm\lambda_j$ captures residual gene-gene dependence beyond the spatial mean structure, inducing the covariance $\mathrm{Cov}(\bm Y(\bm{s}_i)\mid \bm B,\bm\alpha,\bm\Lambda,\bm\Psi) = \bm\Lambda\bm\Lambda^\top+\bm\Psi$, where $\bm\Lambda=(\bm\lambda_1,\dots,\bm\lambda_p)^\top$ is the loading matrix and $\bm\Psi=\mathrm{diag}(\psi_1,\dots,\psi_p)$. 
In matrix notation, let $\bm Y$ be the $n\times p$ matrix with $(i,j)^{\text{th}}$ entry $Y_j(\bm{s}_i)$, $\bm X$ be the $n\times K$ basis matrix with $i^{\text{th}}$ row $\bm X(\bm{s}_i)^\top$, $\bm B=(\bm\beta_1,\dots,\bm\beta_p)$, $\bm F=(\bm f_1,\dots,\bm f_n)^\top$ and $\bm E$ be the $n\times p$ matrix with $(i,j)^{\text{th}}$ entry $\epsilon_{ij}$. Compactly, we write:
\begin{equation}
\label{matrix-low-rank-model}
\bm Y = \bm 1_n\bm\alpha^\top + \bm X\bm B + \bm F\bm\Lambda^\top + \bm E.
\end{equation}
Our methodology does not require additional assumptions on the spatial basis functions $X_k(\cdot)$. However, throughout this article, including both simulations and real data analyses, we use cubic B-spline basis functions \citep{de1978practical}. Explicit expressions for these basis functions are provided in Supplementary Materials Section~\ref{supp:cubic-b-spline}.

\subsection{Spatially Varying Gene Selection via Spike-and-Slab Prior}
\label{sub:spike-slab}

The goal of JASPER is to identify SVGs. We introduce binary indicators $\bm\gamma(\gamma_1,\dots,\gamma_p)^\top$, where $\gamma_j=1$ indicates that gene $j$ is spatially varying and $\gamma_j=0$ otherwise. Conditional on $\gamma_j$, we place the spike-and-slab prior over the spatial basis coefficients $\bm \beta_j$:
\begin{equation}
\label{eq:beta-prior}
\bm\beta_j\mid \gamma_j,g \sim
(1-\gamma_j)\delta_{\bm 0_K} + \gamma_j \mathcal N_K\left\{\bm 0_K, g(\bm X^\top\bm X)^{-1} \right\},
\end{equation}
where $g$ controls the global slab variance for spatial effects. Thus, $\gamma_j=0$ forces $\bm\beta_j=\bm 0_K$, whereas $\gamma_j=1$ allows gene $j$ to have a nonzero spatial basis representation. We use a Beta-Bernoulli prior for the indicators, i.e., $\gamma_j\mid \delta \overset{IID}{\sim} \mathrm{Bernoulli}(\delta)$ with $\delta\sim \mathrm{Beta}(c,d)$, yielding the following marginal for $\bm\gamma$:
\begin{equation}
\label{eq:gamma-prior}
f(\bm\gamma) =
\frac{\Gamma(c+d)}{\Gamma(c)\Gamma(d)}
\frac{\Gamma(c+q)\Gamma(d+p-q)}{\Gamma(c+d+p)},
\qquad
q=\sum_{j=1}^p\gamma_j.
\end{equation}
Finally, we complete the specification of our model by choosing the following priors on the remaining model components:
\begin{align}
\label{eq:other-priors}
\begin{gathered}
\alpha_j \sim \mathcal N(0,\sigma_\alpha^2),\qquad
\bm f_i \sim \mathcal N_r(\bm 0_r,\bm I_r),\qquad
\bm\lambda_j \sim \mathcal N_r(\bm 0_r,\tau_\lambda^2\bm I_r),\\
\psi_j \sim \mathcal{IG}(a_\psi,b_\psi),\qquad
\phi_j \sim \mathrm{Gamma}(a_\phi,b_\phi),\qquad
g \sim \mathcal{IG}(a_g,b_g).
\end{gathered}
\end{align}
%

Defining $\bm \Theta = (\bm Y,\bm\alpha,\bm B,\bm F,\bm\Lambda,\bm\psi,\bm\phi,g,\bm\gamma)$, the full JASPER model is:
\begin{multline}
\label{likelihood-prior-dependence}
f(\bm C,\bm \Theta)
=
\underbrace{f(\bm C\mid \bm Y,\bm\phi)}_{\text{Counts model}}
\underbrace{f(\bm Y\mid \bm\alpha,\bm B,\bm F,\bm\Lambda,\bm\psi)}_{\text{Spatial regression}}
\underbrace{f(\bm B\mid \bm\gamma,g)f(\bm\gamma)}_{\text{SVG model}}
\underbrace{f(\bm F, \bm\Lambda, \bm\alpha, \bm\psi, \bm\phi, g)}_{\text{Other priors}}.
\end{multline}

\subsection{Posterior Inference via data-augmented Gibbs sampler}
\label{sub:full-conditionals}

Posterior inference for JASPER is performed using an efficient Gibbs sampler. To tackle the intractable negative-binomial likelihood we the Polya-Gamma data augmentation~\citep{polyagamma} to sample the latent expression variables $Y_j(\bm{s}_i)$.
The parameters $\bm f_i$, $\bm\lambda_j$, $\alpha_j$, $\psi_j$, and $g$ have conjugate updates. Post marginalization of $\bm \beta_j$, the SVG indicators $\gamma_j$ are updated using a collapsed Gibbs step. Conditional on $\gamma_j=1$, $\bm\beta_j$ is sampled from its Gaussian full conditional, while $\bm\beta_j$ is fixed at $\bm 0_K$ when $\gamma_j=0$. Finally, $\phi_j$'s are updated using a log-scale random-walk Metropolis-Hastings step. Full conditional distributions and derivations are provided in Supplementary Materials Section~\ref{supp:jasper-full-conditionals}.

\subsection{Selection of Spatially Varying Genes}

The posterior samples of $\bm\gamma$ are used to estimate the posterior probability of inclusion for each gene, i.e., $\mathrm{PPI}_j = \tfrac1M \sum_{m=1}^M \gamma_j^{(m)}.$ We select gene $j$ as SVG if $\mathrm{PPI}_j\ge t$, where the threshold $t$ is chosen to control the posterior expected false discovery rate~\citep{muller2004optimal}:
$$\mathrm{peFDR}(t)
=\frac{\sum_{j=1}^p(1-\mathrm{PPI}_j)\mathbb I(\mathrm{PPI}_j\ge t)}{\sum_{j=1}^p \mathbb I(\mathrm{PPI}_j\ge t)},$$
i.e., we choose the smallest threshold $t$ such that $\mathrm{peFDR}(t)\le 0.05$ to select SVGs.


\section{Simulation Study}
\label{section:simulations}

For performance evaluation of JASPER, we conducted three simulation studies designed to assess information sharing across correlated genes, robustness to spatial kernel misspecification, and detection accuracy under varying levels of SVG sparsity. Across all settings, JASPER consistently outperformed existing methods, particularly in the presence of strong gene-gene dependence and complex spatial structure.

\subsection{Simulations for Co-Expression Modeling}
\label{sub:simulations-coexp}

\begin{figure}
     \centering
     \begin{subfigure}[b]{\textwidth}
    \includegraphics[width=\textwidth]{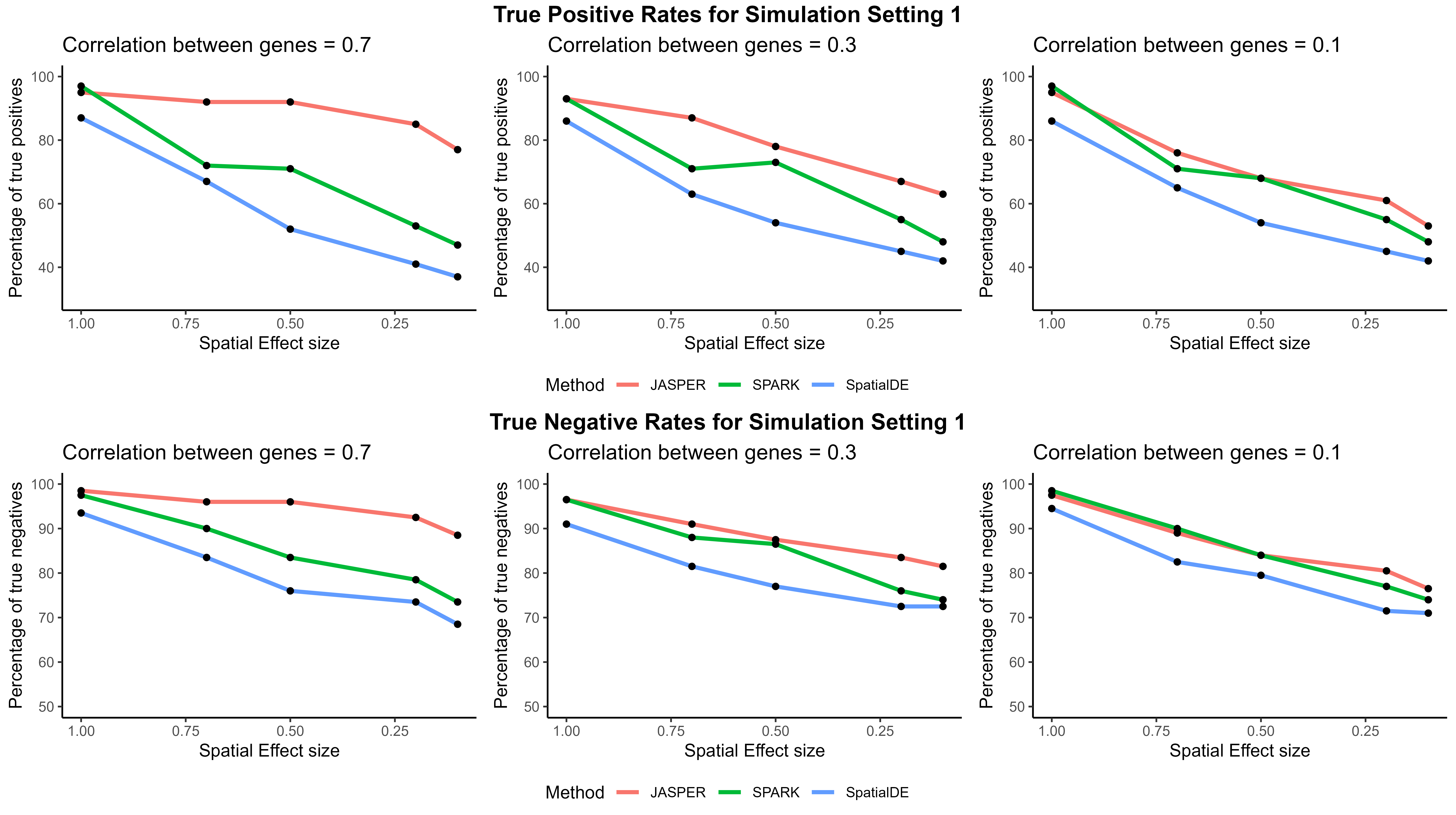}
     \end{subfigure}
     \hfill
     \begin{subfigure}[b]{\textwidth}
    \includegraphics[width=\textwidth]{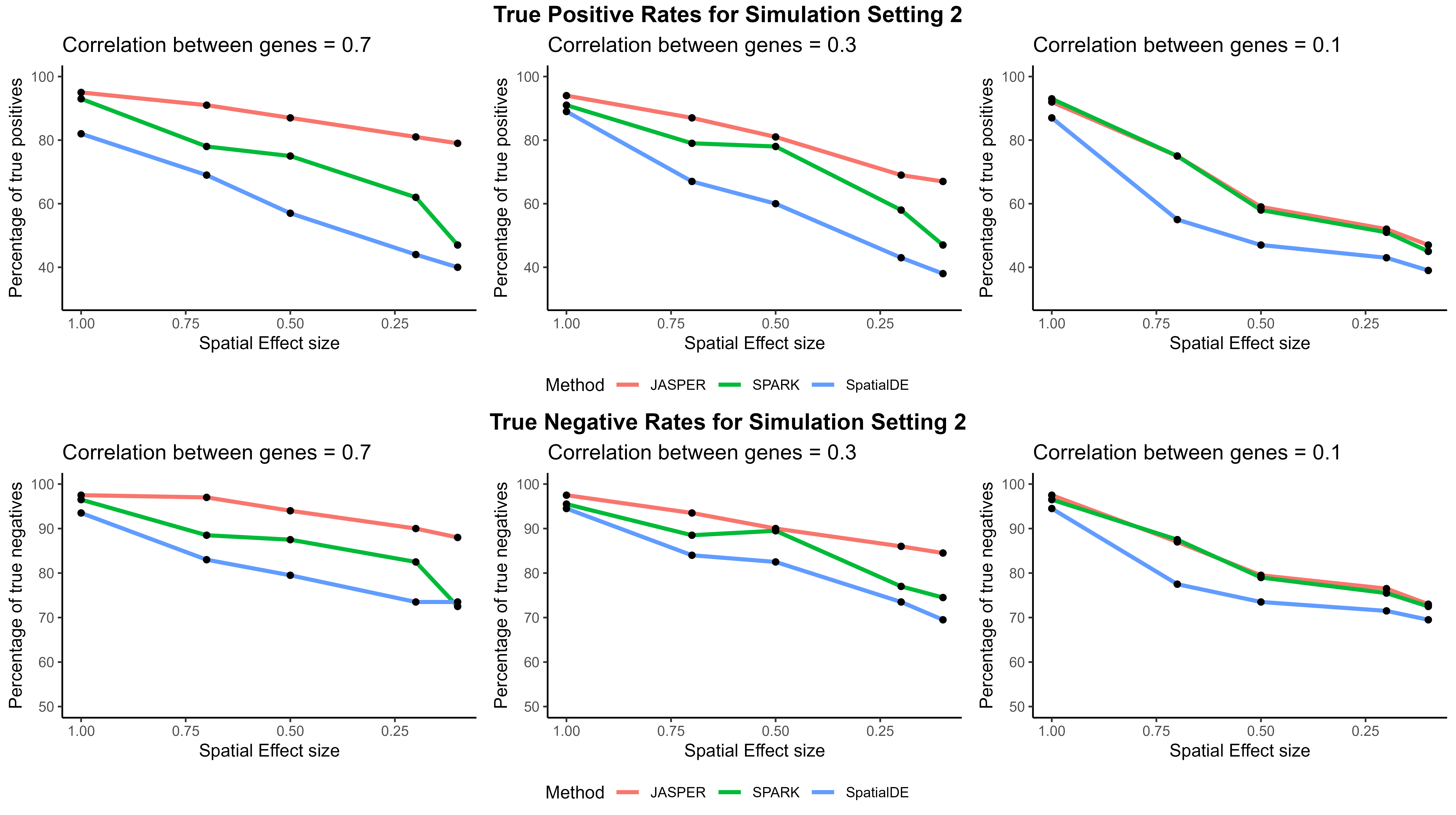}
     \end{subfigure}
     \captionsetup{font={stretch=1.1}}
         \caption{Simulation-based comparison of JASPER, SPARK, and SpatialDE in terms of percentage of true positives and negatives for two simulation schemes for different correlation levels and degrees of spatial dependence among genes based on (1) A model close to the underlying model of JASPER; and (2) A model based on dependent Gaussian processes.}
        \label{fig:sims}
\end{figure}

We considered two simulation models with dependent gene expressions: one closely resembled JASPER, and the other with multivariate Gaussian processes (GPs). Simulation details are in Table~\ref{tab:simulation-settings} of Supplementary Materials Section~\ref{supp:sim-exps}.
Each simulation model has two parameters $0 \leq \mathfrak{A}, \rho \leq 1$, where $\mathfrak{A}$ (spatial effect size) controls the strength of spatial association (high value implies more spatial variation), while $\rho$ controls the correlation between pairs of simulated genes (high implies more correlated). 
We chose $\mathfrak{A} \in \{1,0.75,0.5,0.2,0.15\}$ and $\rho \in \{0.7, 0.3, 0.1\}$. For each combination, we chose $n = 200$ locations uniformly from the unit square ($[0,1] \times [0,1]$), and $p=100$ simulated genes from each location, of which $20$ were randomly chosen to be SVGs. These simulated datasets were fitted with JASPER, along with competing models: SPARK \citep{spark} and SpatialDE \citep{spatialde}. Figure \ref{fig:sims} plots the true positive and negative rates (sensitivity and specificity) over $5$ repetitions.

Figure~\ref{fig:sims} shows that all three methods perform well for strong spatial effects, but JASPER increasingly outperforms SPARK and SpatialDE as spatial effects weaken. This advantage is most pronounced under stronger gene-gene correlation, demonstrating the benefit of joint modeling for SVG detection.

\subsection{Simulations for Kernel Independence of JASPER}
\label{sub:simulations-kernel}

This simulation study demonstrates the effect of kernel misspecification on SVG detection. We simulate the normalized gene expressions for 50 genes at 200 randomly chosen spatial locations in the unit square, where for gene $j$ at location $\bm{s}_i$ $Y_j(\bm{s}_i) = f_j(\bm{s}_i) + \epsilon_{ij}$, with $f_j \sim GP(0, K_j(\cdot, \cdot))$ for $1 \leq j \leq 10$,  $f_j(x) = 0$ for $11 \leq j \leq 50$, $\epsilon \sim N(0,0.1)$, and $K_j(\bm{s}_i, \bm{s}_k) = \sigma^2 \exp(-2sin^2(\pi d(\bm{s}_i, \bm{s}_k)/p)/l_j^2)$ with $\sigma^2 = 0.1$, $p = 0.25$, and $l_j \sim \mathrm{Gamma}(5,2)$. We applied SpatialDE, SOMDE \citep{somde}, and JASPER to this data across $50$ repetitions. The true positive rates were $43.2\% (\pm 8.8\%)$ for SpatialDE, $63.5\% (\pm 7.3\%)$ for SOMDE and $93.5\% (\pm 4.1\%)$ for JASPER, which shows that if the spatial expression patterns do not conform well with the respective covariance kernel, a significant number of SVGs will not be detected by kernel-based methods, while JASPER can easily adapt to a wider variety of expression patterns.

\subsection{Simulations for Different SVG Sparsity}

While existing methods check each gene one-by-one for spatial variability, JASPER does simultaneous SVG detection. Hence the SVG count in a given data may affect SVG detection. To gauge JASPER's performance for different SVG sparsity levels, we conduct a simulation study with varying proportion of SVGs. We use the GP-based simulation settings described in Section \ref{sub:simulations-coexp}. We fix $\rho = 0.3$ and $\mathfrak{A} = 0.2$, and randomly select $n = 200$ locations from the unit square as before for $p=100$ genes. The SVG counts are varied as $\{20, 40, 60\}$. For each count, the true positive and negative rates are averaged over $50$ repetitions. The average true positive rates were 73.52\%, 87.21\% and 97.3\% respectively while the true negative rates were 83.92\%, 86.72\% and 88.52\%, respectively. These results conclude that with increasing proportion of SVGs, the detection problem for JASPER becomes more accurate particularly in terms of true positive rates. This improvement can probably be attributed to correlation between genes, allowing JASPER to borrow information across SVGs to for better learning the underlying correlation pattern. 


\section{Real Data Applications}
\label{section:real-data}

We now demonstrate the performance of JASPER in terms of statistical power and biological interpretability compared to two state-of-the-art competing methods, SPARK \citep{spark} and SpatialDE \citep{spatialde} for the SRT datasets mentioned in Section \ref{section:data-description}. 

\subsection{Application to Human Breast Cancer dataset}

We apply JASPER to the motivating human breast cancer dataset from the iIMPACT study \citep{iimpact}. As a preprocessing step, we filtered out practically unobserved genes, i.e., genes with fewer than 100 total counts over all locations. Additionally, we eliminated genes with very low spatial signal using a Bayes-factor based screening method; see Supplementary Materials Section~\ref{sub:simplified-versions}. Post filtering, we ran the Gibbs sampler for 5000 iterations with the first 3000 as burn-in. For comparison, we ran SPARK and SpatialDE, using their default FDR threshold of 5\%.

JASPER detected 672 SVGs, and SPARK and SpatialDE detected 567 and 709 SVGs, respectively. We used three different modes of comparison to evaluate the biological significance of the detected SVGs from each method. 

First, we compared the overlap between the SVGs detected by each method with known breast cancer gene sets from two independent gene databases, namely, the Harmonizome database \citep{harmonizome} and the Catalogue Of Somatic Mutations In Cancer (COSMIC) database \citep{cosmic}. As reported in Table \ref{tab:breast-cancer}, the SVGs detected by JASPER have greater overlap with both biologically curated gene sets as compared to SPARK and SpatialDE. Comparing the genes uniquely detected by each method, we again look at the overlap of the uniquely detected genes with the two databases. Additionally, we calculated the Moran's I statistic for these uniquely detected genes to quantitatively compare the degree of spatial autocorrelation of gene expression. As reported in Table \ref{tab:breast-cancer}, on average the Moran's I statistic for the genes uniquely detected by JASPER (0.61) is higher than those detected by SPARK (0.41) and SpatialDE (0.34). Genes uniquely identified by JASPER include COX6C, a known \citep{ref-cox6c} indicator of hormone-responsive breast cancer and EFNA1, a gene known to be involved in tumor cell proliferation, invasiveness, and metastasis \citep{ref-efna1}.
\begin{figure}[!htp]
\centering
\captionsetup{font={stretch=1.1}}
\includegraphics[width=\textwidth]{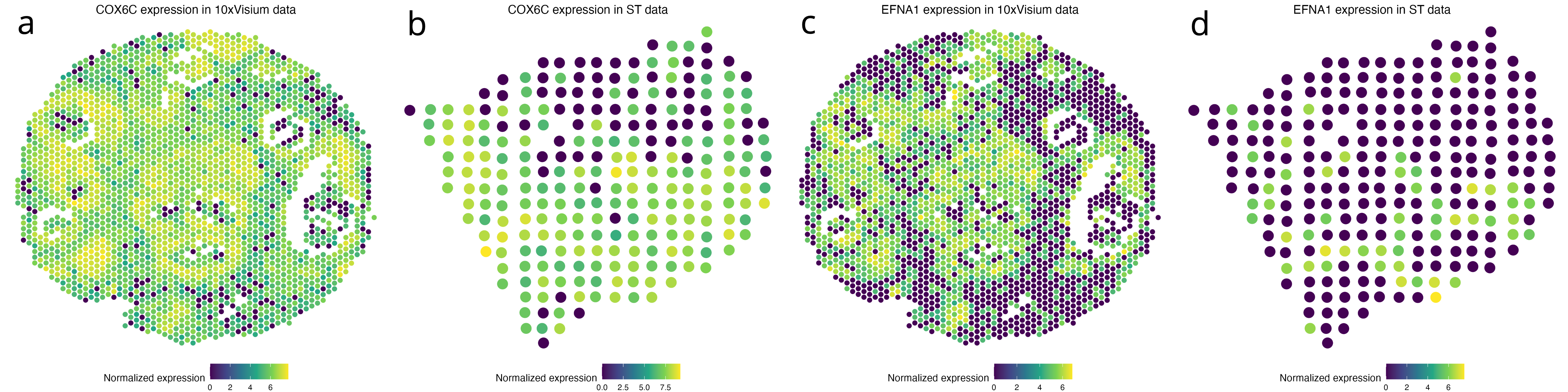}
\caption{Normalized spatial expression of genes COX6C (a-b) and EFNA1 (c - d) in two human breast cancer datasets. Both these genes were detected uniquely by JASPER in the 10x Visium dataset and were also detected in the ST dataset.}
\label{fig:bc-expression}
\end{figure}

Second, we compared the biological consistency of the detected SVGs by applying all three methods to another publicly available human breast cancer dataset from an earlier study \citep{spatialtranscriptomics}. This dataset, consisting of 5,260 genes and 250 locations, was obtained using the older ST technology, which has a lower spatial resolution compared to the 10x Visium technology. Nevertheless, we found significant overlap between the genes detected by JASPER in the two datasets. More specifically, 89.8\% of the genes detected by JASPER in the ST breast cancer dataset are also detected in the more recent Visium data. This proportion is much lower for SPARK (61.3\%) and SpatialDE (54.9\%). The aforementioned genes COX6C and EFNA1, were detected by JASPER in both datasets. The normalized expression plots of these genes in both datasets are provided in Figure \ref{fig:bc-expression}, which have clear spatial patterns.

Third, a comprehensive gene set enrichment analysis (GSEA) was conducted for SVGs identified by all three methods using g:Profiler \citep{Raudvere2019}. A total of 1,527 Gene Ontology (GO) terms were enriched in the SVGs identified by JASPER, significantly outperforming SPARK and SpatialDE, which identified only 867 and 87 enriched GO terms, respectively. Further, using the genes exclusively identified by JASPER, the top enriched molecular functions were related to ``cell adhesion molecule binding" (GO:0050839; $p_{adj} = 8.84 \times 10^{-20}$) and ``extracellular matrix structural constituent" (GO:0005201; $p_{adj} = 1.17 \times 10^{-17}$). These highly statistically significant terms are also highly biologically relevant to breast cancer in general and ductal carcinomas in particular. Notably, cell adhesion and extracellular matrix (ECM) interactions play a key role in the progression of ductal carcinomas where tumor cells often exploit these pathways to enhance metastatic potential \citep{callahan2011human, fang2009gata}. Additionally, enriched biological processes such as ``extracellular matrix organization" (GO:0030198) and ``immune response" (GO:0006955) suggest that the uniquely identified genes are involved in modulating the tumor microenvironment, influencing immune cell interactions, and contributing to the chronic inflammatory state often associated with tumorigenesis \citep{Kufe2009, hurtado2011foxa1}. 

Lastly, pathway analysis based on the JASPER-detected SVGs using the Kyoto Encyclopedia of Genes and Genomes (KEGG) and Reactome databases provided further insights into the molecular mechanisms underlying breast cancer. Enrichment in pathways such as ``ECM-receptor interaction" (KEGG:04512) and ``Signaling by Interleukins" (REAC:R-HSA-449147) underscores the importance of ECM dynamics and cytokine signaling in cancer biology. The ``ECM-receptor interaction" pathway is known to facilitate cancer cell migration and invasion by modulating integrin signaling, while ``Signaling by Interleukins" plays a critical role in maintaining the pro-inflammatory environment that supports tumor growth and immune evasion \citep{Ferrara2002, Kristiansen2003, Satelli2011}. Some key genes detected by JASPER include: \textit{ERBB2}, which is often amplified in aggressive breast cancers and contribute to signaling pathways that promote cell proliferation and survival, \textit{MUC1}, a glycoprotein overexpressed in breast cancer  and involved in cell signaling and adhesion, and \textit{GATA3}, a transcription factor which regulates estrogen receptor signaling in luminal breast cancer subtypes. These biological findings using the gene set uniquely identified by JASPER not only underscores its robustness in detecting genes integral to the underlying biological mechanisms of breast cancer, but may also potentially contribute in identifying important therapeutic targets and biomarkers, potentially enhancing personalized treatment strategies for breast cancer patients.

\begin{table}[!ht]
\centering
\captionsetup{font={stretch=1.1}}
\caption{Comparison of SVG detection methods for the human breast cancer dataset generated using 10x Visium technology.}
\label{tab:breast-cancer}

\small

\begin{tabular}{llccc}
\toprule
\textbf{Category} & \textbf{Metric} & \textbf{JASPER} & \textbf{SPARK} & \textbf{SpatialDE} \\
\midrule

\multirow{3}{*}{SVGs}
& SVGs detected          & 672 & 567 & 709 \\
& Harmonizome overlap    & 207 & 134 & 37  \\
& COSMIC overlap         & 118 & 95  & 91  \\

\midrule

\multirow{4}{*}{Unique SVGs}
& Unique SVGs            & 234 & 195 & 236 \\
& Harmonizome overlap    & 56  & 37  & 15  \\
& COSMIC overlap         & 29  & 29  & 13  \\
& Average Moran's $I$    & 0.61 & 0.41 & 0.34 \\

\midrule

\multirow{3}{*}{Replication Analysis}
& SVGs in secondary data & 342 & 290 & 113 \\
& Replicated SVGs        & 307 & 178 & 62  \\
& Replication rate       & (89.8\%) & (61.3\%) & (54.9\%) \\

\bottomrule
\end{tabular}
\end{table}

\subsection{Application to Mouse visual cortex STARmap data}

\begin{figure}[!ht]
    \centering
    \includegraphics[width = \textwidth]{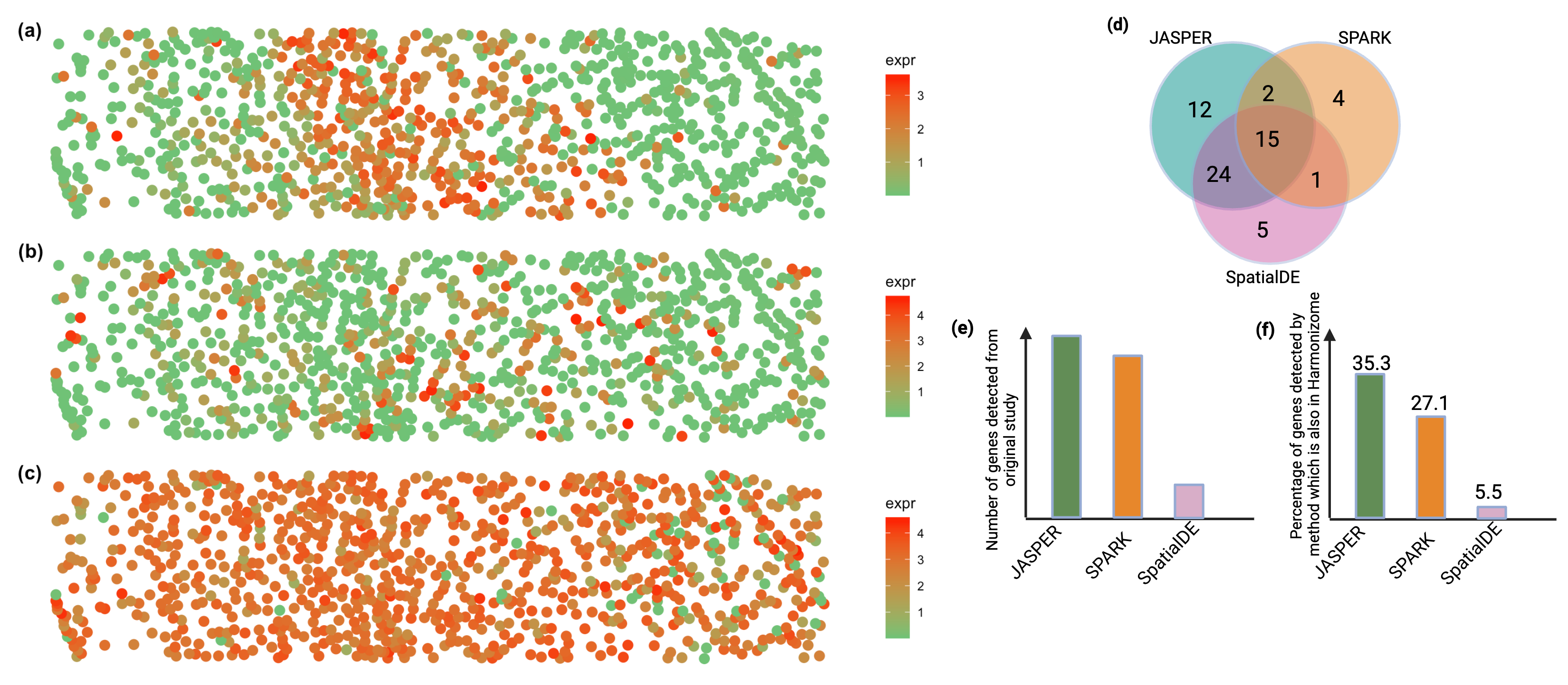}
    \captionsetup{font={stretch=1.1}}
     \caption{(a)-(c) Expression patterns of three genes from the STARmap dataset.  (a) Banded spatial pattern; (b) Sparse collection of clusters of high expression; (c) No spatial pattern; (d) Diagram showing distribution of genes detected by JASPER, SPARK and SpatialDE, (e) Number of genes from the original study detected by each method. (f) Percentage of genes detected by each method which are biologically significant according to the Harmonizome database.}
    \label{fig:starmap-patterns}
\end{figure}

The STARmap data \citep{starmap} measures gene expression in the mouse visual cortex after one hour of light exposure following four days of dark exposure, to study how light exposure modulates gene expression and how these effects vary spatially across the visual cortex. We analyzed one randomly selected mouse, comprising $160$ genes measured across $891$ spatial locations. Representative spatial expression patterns are shown in Figure~\ref{fig:starmap-patterns}.

The same preprocessing steps were performed under settings mentioned in the breast cancer data application for each SVG detection method. JASPER detected 53 SVGs of various spatial patterns, compared to 45 and 22 SVGs detected by SPARK and SpatialDE, respectively. Most of the genes detected to be spatial by JASPER were indeed highly spatially varying. Figure \ref{fig:starmap-patterns} (a) \& (b) displays the spatial patterns for two such genes. Further, JASPER captured all 13 SVGs  highlighted originally in \cite{starmap} compared to 11 detected by SPARK and only 3 by SpatialDE. As these 13 genes are known to be biologically important and also known to exhibit spatial dependencies, this result provides an empirical validation of JASPER against competing methods. Among the JASPER-detected non-SVGs, most of them had low or zero expression values with only a few non-zero values. In contrast, SPARK and SpatialDE were often unable to capture genes that had sparse clusters of high expression. For example, in Figure \ref{fig:starmap-patterns}(b), gene Gad1, which encodes the enzyme glutamate decarboxylase 67 and is crucial for the development and function of the nervous system including spatial memory, was only detected by JASPER. Selected JASPER-detected SVGs along with their biological significance are provided in Table \ref{genes-starmap} of Supplementary Materials. 
Lastly, as an additional validation of the genes identified by these methods, we compared the SVGs identified by each of the methods with the curated list of 898 genes associated with visual areas from the Harmonizome database \citep{harmonizome}. As shown in Figure \ref{fig:starmap-patterns}(f), 35.3\% of the genes obtained by JASPER were part of the Harmonizome list, compared to 27.1\% and 5.5\% for SPARK and SpatialDE respectively, which demonstrates that JASPER is able to capture more genes associated with the visual areas. 

\subsection{Mouse Olfactory Bulb Dataset}
\label{sub:mousebulbdata}

\begin{figure}[!ht]
    \centering
    \includegraphics[width = \textwidth]{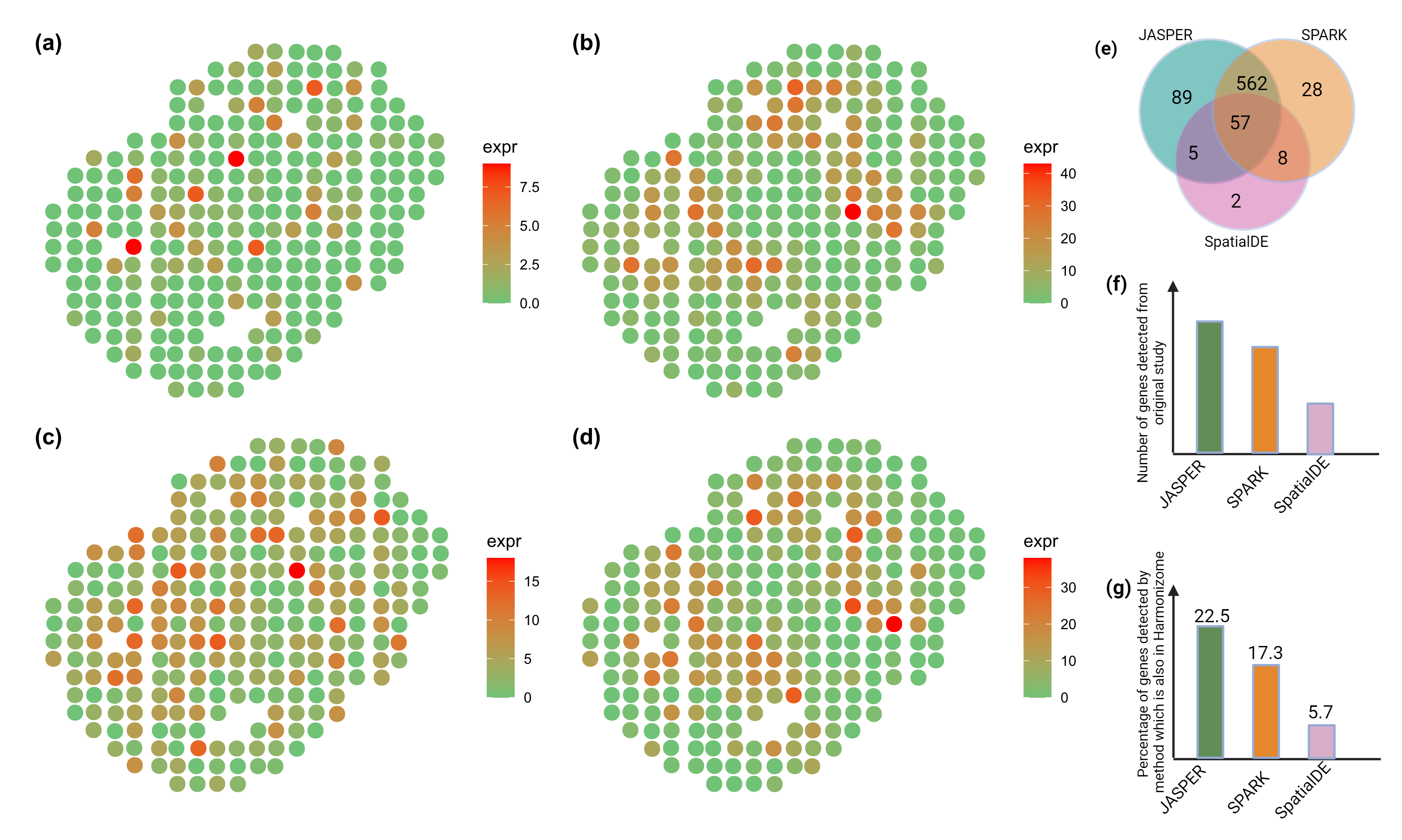}
    \captionsetup{font={stretch=1.1}}
    \caption{ (a) - (d) Some expression patterns of different genes from the Mouse olfactory bulb dataset.  (a) Detected by SPARK only; (b) Detected by SpatialDE only; (c) Detected by all three methods; (d) Detected by JASPER only; (e) Diagram showing distribution of genes detected by JASPER, SPARK and SpatialDE, (f) Number of genes from the original study detected by each method; (g) Percentage of genes that are of biological relevance according to the Harmonizome database.}
    \label{fig:olfactory-patterns}
\end{figure}

The mouse olfactory bulb dataset \citep{spatialtranscriptomics} contains expression measurements for 16,573 genes across 265 spots. SPARK and SpatialDE identified 655 and 72 SVGs, respectively, whereas JASPER identified 713. Representative expression profiles for genes detected by each method are shown in Figure~\ref{fig:olfactory-patterns}(a,b,d). Validation against the original study \citep{spatialtranscriptomics} is shown in Figure~\ref{fig:olfactory-patterns}(f). Most genes uniquely detected by JASPER exhibit clear spatial expression patterns. Notably, JASPER recovered all 10 genes highlighted in the original study, compared with 8 for SPARK and 4 for SpatialDE. We further compared the SVGs from each method with 593 genes associated with olfactory regions of the mouse brain in the Harmonizome database. JASPER achieved the highest overlap, with 22.5\% of its detected SVGs present in Harmonizome, compared with 17.3\% for SPARK and 5.7\% for SpatialDE; see Figure~\ref{fig:olfactory-patterns}(g). A selection of SVGs detected only by JASPER, along with their biological significance, is provided in Supplementary Table~\ref{genes-MOB}. These results further support the improved power of JASPER for identifying biologically relevant SVGs in olfactory tissue.

\section{Discussion}
\label{section:discussion}

We presented JASPER, a fully Bayesian model for detecting joint detection of SVGs in SRT datasets. Unlike most existing SVG detection methods, JASPER is designed to model multiple genes simultaneously to account for co-expression. Additionally, the use of spatial basis functions instead of pre-specified Gaussian process kernels is a key feature that allows us to target a wide variety of spatial patterns without making implicit assumptions (for example stationarity) about the patterns in question. 

Through our simulations, we demonstrated that JASPER is able to achieve higher selection accuracy compared to state-of-the-art competing methods by leveraging shared information across genes. Our application of the method to real datasets confirmed that our model was indeed able to capture a higher number of biologically significant genes (according to the respective original studies and as well as an independent publicly available gene database) compared to the competing methods.

There are several possible future research directions for JASPER. Spatial variability in gene expression can often be explained by the presence of different cell types in various regions of the tissue \citep{yu2022identification}. A straightforward extension of JASPER could involve modeling gene expression separately for each cell type. Another extension could be through multi-omic integration of gene expressions with other spatial omic data, which may potentially provide a more context-based methodology for SVG identification. Finally, genes can be clustered by their expression patterns, which could be incorporated into the structure of the co-expression covariance matrix.

\section*{Data and Code Availability}

Data used in this work are publicly available at \href{https://lce.biohpc.swmed.edu/star/index.html}{lce.biohpc.swmed.edu/star}. A \texttt{Python} implementation of JASPER can be found at \href{https://github.com/pritamdey/pyJASPER}{github.com/pritamdey/pyJASPER}.

\section*{AI use disclosure}

Generative AI tools were used to assist with language editing and drafting limited portions of the biological interpretation. The authors verified and take full responsibility for all analyses, conclusions, and references.

\bibliographystyle{unsrtnat}
\bibliography{references}

\newpage

\appendix
\thispagestyle{empty}
\spacingset{1.3}

\begin{center}
\hrule height 1.5pt
\vspace{2em}
{\LARGE\textbf{Appendices}}
\end{center}
\noindent
\hrule height 1pt
\vspace{1.5em}

\newcommand{\beginproofs}
{
\renewcommand{\thesection}{\Alph{section}}
\renewcommand{\thetable}{S\arabic{table}}
\renewcommand{\thefigure}{S\arabic{figure}}
\renewcommand{\theequation}{S\arabic{equation}}
}
\beginproofs

\section{Full Conditional Distributions for the JASPER Gibbs Sampler}
\label{supp:jasper-full-conditionals}

We derive the full conditional distributions used in the Gibbs sampler for JASPER. Recall that the model is
\[
C_j(\bm{s}_i)\mid Y_j(\bm{s}_i),\phi_j
\sim
\mathrm{NB}\{\mu_j(\bm{s}_i),\phi_j\},
\qquad
\mu_j(\bm{s}_i)=N_i\exp\{Y_j(\bm{s}_i)\},
\]
and
\[
Y_j(\bm{s}_i)
=
\alpha_j
+
\bm X(\bm{s}_i)^\top \bm\beta_j
+
\bm f_i^\top\bm\lambda_j
+
\epsilon_{ij},
\qquad
\epsilon_{ij}\sim \mathcal N(0,\psi_j).
\]
The priors are
\[
\bm f_i\sim \mathcal N_r(\bm 0_r,\bm I_r),
\qquad
\bm\lambda_j\sim \mathcal N_r(\bm 0_r,\tau_\lambda^2\bm I_r),
\qquad
\alpha_j\sim \mathcal N(0,\sigma_\alpha^2),
\]
\[
\psi_j\sim \mathcal{IG}(a_\psi,b_\psi),
\qquad
\phi_j\sim \mathrm{Gamma}(a_\phi,b_\phi),
\qquad
g\sim \mathcal{IG}(a_g,b_g),
\]
and
\[
\bm\beta_j\mid \gamma_j,g
\sim
(1-\gamma_j)\delta_{\bm 0_K}
+
\gamma_j
\mathcal N_K\left\{\bm 0_K,g(\bm X^\top\bm X)^{-1}\right\}.
\]

\subsection{Polya-Gamma Augmentation}

For the negative binomial likelihood, define
\[
\eta_{ij}
=
Y_j(\bm{s}_i)+\log N_i-\log\phi_j.
\]
Using $\mu_j(\bm{s}_i)=N_i\exp\{Y_j(\bm{s}_i)\}$, the negative binomial likelihood can be written, up to terms not involving $\eta_{ij}$, as
\[
p\{C_j(\bm{s}_i)\mid Y_j(\bm{s}_i),\phi_j\}
\propto
\frac{\exp(\kappa_{ij}\eta_{ij})}
{\{1+\exp(\eta_{ij})\}^{C_j(\bm{s}_i)+\phi_j}},
\]
where
\[
\kappa_{ij}
=
\frac{C_j(\bm{s}_i)-\phi_j}{2}.
\]
By the Polya-Gamma identity,
\[
\frac{\exp(\kappa_{ij}\eta_{ij})}
{\{1+\exp(\eta_{ij})\}^{C_j(\bm{s}_i)+\phi_j}}
\propto
\int_0^\infty
\exp\left\{
-\frac{\omega_{ij}\eta_{ij}^2}{2}
+
\kappa_{ij}\eta_{ij}
\right\}
p(\omega_{ij})\,d\omega_{ij},
\]
where
\[
\omega_{ij}\mid -
\sim
\mathrm{PG}\{C_j(\bm{s}_i)+\phi_j,\eta_{ij}\}.
\]
Conditional on $\omega_{ij}$, the likelihood contribution for $Y_j(\bm{s}_i)$ is Gaussian. Completing the square gives
\[
-\frac{\omega_{ij}}{2}\eta_{ij}^2+\kappa_{ij}\eta_{ij}
=
-\frac{\omega_{ij}}{2}
\left[
Y_j(\bm{s}_i)-z_{ij}
\right]^2
+
\text{constant},
\]
where
\[
z_{ij}
=
\frac{\kappa_{ij}}{\omega_{ij}}
-
\log N_i
+
\log\phi_j.
\]
Therefore,
\[
z_{ij}\mid Y_j(\bm{s}_i),\omega_{ij}
\sim
\mathcal N\left\{Y_j(\bm{s}_i),\omega_{ij}^{-1}\right\}.
\]

\subsection{Full Conditional for \texorpdfstring{$Y_j(\bm{s}_i)$}{Yj(si)}}

Let
\[
m_{ij}
=
\alpha_j
+
\bm X(\bm{s}_i)^\top\bm\beta_j
+
\bm f_i^\top\bm\lambda_j.
\]
The prior model implies
\[
Y_j(\bm{s}_i)\mid -
\sim
\mathcal N(m_{ij},\psi_j),
\]
while the Polya-Gamma pseudo-likelihood gives
\[
z_{ij}\mid Y_j(\bm{s}_i),\omega_{ij}
\sim
\mathcal N\{Y_j(\bm{s}_i),\omega_{ij}^{-1}\}.
\]
Thus,
\[
p\{Y_j(\bm{s}_i)\mid -\}
\propto
\exp\left[
-\frac{\omega_{ij}}{2}
\{z_{ij}-Y_j(\bm{s}_i)\}^2
-
\frac{1}{2\psi_j}
\{Y_j(\bm{s}_i)-m_{ij}\}^2
\right].
\]
Completing the square,
\[
Y_j(\bm{s}_i)\mid -
\sim
\mathcal N(\widetilde m_{ij},\widetilde v_{ij}),
\]
where
\[
\widetilde v_{ij}
=
\left(\omega_{ij}+\psi_j^{-1}\right)^{-1},
\qquad
\widetilde m_{ij}
=
\widetilde v_{ij}
\left(
\omega_{ij}z_{ij}
+
\psi_j^{-1}m_{ij}
\right).
\]

\subsection{Full Conditional for Factor Scores $f_i$}

For location $i$, define
\[
\bm y_i=\bm Y(\bm{s}_i),
\qquad
\bm a_i
=
\bm\alpha+\bm B^\top\bm X(\bm{s}_i),
\qquad
\bm\Psi=\mathrm{diag}(\psi_1,\dots,\psi_p).
\]
Then
\[
\bm y_i\mid \bm f_i,-
\sim
\mathcal N_p(\bm a_i+\bm\Lambda\bm f_i,\bm\Psi),
\qquad
\bm f_i\sim \mathcal N_r(\bm 0_r,\bm I_r).
\]
Therefore,
\[
p(\bm f_i\mid -)
\propto
\exp\left[
-\frac12
(\bm y_i-\bm a_i-\bm\Lambda\bm f_i)^\top
\bm\Psi^{-1}
(\bm y_i-\bm a_i-\bm\Lambda\bm f_i)
-
\frac12\bm f_i^\top\bm f_i
\right].
\]
Collecting terms in $\bm f_i$ gives
\[
\bm f_i\mid -
\sim
\mathcal N_r(\bm m_{f_i},\bm V_f),
\]
where
\[
\bm V_f
=
\left(
\bm I_r+\bm\Lambda^\top\bm\Psi^{-1}\bm\Lambda
\right)^{-1},
\]
and
\[
\bm m_{f_i}
=
\bm V_f
\bm\Lambda^\top\bm\Psi^{-1}
\left\{
\bm Y(\bm{s}_i)-\bm\alpha-\bm B^\top\bm X(\bm{s}_i)
\right\}.
\]

\subsection{Full Conditional for Factor Loadings $\lambda_j$}

For gene $j$, define
\[
\bm r_{\lambda,j}
=
\bm Y_j-\alpha_j\bm 1_n-\bm X\bm\beta_j.
\]
Then
\[
\bm r_{\lambda,j}\mid \bm\lambda_j,-
\sim
\mathcal N_n(\bm F\bm\lambda_j,\psi_j\bm I_n),
\qquad
\bm\lambda_j\sim \mathcal N_r(\bm 0_r,\tau_\lambda^2\bm I_r).
\]
Thus,
\[
p(\bm\lambda_j\mid -)
\propto
\exp\left[
-\frac{1}{2\psi_j}
(\bm r_{\lambda,j}-\bm F\bm\lambda_j)^\top
(\bm r_{\lambda,j}-\bm F\bm\lambda_j)
-
\frac{1}{2\tau_\lambda^2}
\bm\lambda_j^\top\bm\lambda_j
\right].
\]
Therefore,
\[
\bm\lambda_j\mid -
\sim
\mathcal N_r(\bm m_{\lambda_j},\bm V_{\lambda_j}),
\]
where
\[
\bm V_{\lambda_j}
=
\left(
\psi_j^{-1}\bm F^\top\bm F+\tau_\lambda^{-2}\bm I_r
\right)^{-1},
\]
and
\[
\bm m_{\lambda_j}
=
\bm V_{\lambda_j}
\psi_j^{-1}
\bm F^\top
\left(
\bm Y_j-\alpha_j\bm 1_n-\bm X\bm\beta_j
\right).
\]

\subsection{Full Conditional for Intercepts $\alpha_j$}

For gene $j$, define
\[
\bm r_{\alpha,j}
=
\bm Y_j-\bm X\bm\beta_j-\bm F\bm\lambda_j.
\]
Then
\[
\bm r_{\alpha,j}\mid \alpha_j,-
\sim
\mathcal N_n(\alpha_j\bm 1_n,\psi_j\bm I_n),
\qquad
\alpha_j\sim \mathcal N(0,\sigma_\alpha^2).
\]
Therefore,
\[
p(\alpha_j\mid -)
\propto
\exp\left[
-\frac{1}{2\psi_j}
(\bm r_{\alpha,j}-\alpha_j\bm 1_n)^\top
(\bm r_{\alpha,j}-\alpha_j\bm 1_n)
-
\frac{\alpha_j^2}{2\sigma_\alpha^2}
\right],
\]
which gives
\[
\alpha_j\mid -
\sim
\mathcal N(m_{\alpha_j},V_{\alpha_j}),
\]
where
\[
V_{\alpha_j}
=
\left(
n\psi_j^{-1}+\sigma_\alpha^{-2}
\right)^{-1},
\]
and
\[
m_{\alpha_j}
=
V_{\alpha_j}
\psi_j^{-1}
\bm 1_n^\top
\left(
\bm Y_j-\bm X\bm\beta_j-\bm F\bm\lambda_j
\right).
\]

\subsection{Collapsed Full Conditional for $\gamma_j$}

For the update of $\gamma_j$, define the residual after removing the non-spatial terms:
\[
\bm r_j
=
\bm Y_j-\alpha_j\bm 1_n-\bm F\bm\lambda_j.
\]
If $\gamma_j=0$, then $\bm\beta_j=\bm 0_K$, and hence
\[
\bm r_j\mid \gamma_j=0,-
\sim
\mathcal N_n(\bm 0_n,\psi_j\bm I_n).
\]
Let the corresponding marginal density be $m_0(\bm r_j)$.

If $\gamma_j=1$, then
\[
\bm r_j\mid \bm\beta_j,\gamma_j=1,-
\sim
\mathcal N_n(\bm X\bm\beta_j,\psi_j\bm I_n),
\]
with
\[
\bm\beta_j\mid \gamma_j=1,g
\sim
\mathcal N_K\left\{\bm 0_K,g(\bm X^\top\bm X)^{-1}\right\}.
\]
Integrating out $\bm\beta_j$ gives
\[
\bm r_j\mid \gamma_j=1,-
\sim
\mathcal N_n
\left[
\bm 0_n,
\psi_j\bm I_n+
g\bm X(\bm X^\top\bm X)^{-1}\bm X^\top
\right].
\]
Let the corresponding marginal density be $m_1(\bm r_j)$.

Because $\delta\sim\mathrm{Beta}(c,d)$ has been integrated out, the prior odds are
\[
\frac{p(\gamma_j=0\mid \bm\gamma_{-j})}
{p(\gamma_j=1\mid \bm\gamma_{-j})}
=
\frac{d+p-1-q_{-j}}{c+q_{-j}},
\qquad
q_{-j}=\sum_{\ell\neq j}\gamma_\ell.
\]
Therefore,
\[
\frac{1-\pi_j}{\pi_j}
=
\frac{m_0(\bm r_j)}{m_1(\bm r_j)}
\frac{d+p-1-q_{-j}}{c+q_{-j}},
\]
and
\[
\gamma_j\mid -
\sim
\mathrm{Bernoulli}(\pi_j).
\]

For computational stability, the marginal likelihood under the active model can be evaluated using the $g$-prior identities. Let $\bm G=\bm X^\top\bm X$ and $\bm P_X=\bm X\bm G^{-1}\bm X^\top$. Then
\[
\log m_0(\bm r_j)
=
-\frac n2\log(2\pi)
-\frac n2\log\psi_j
-\frac{1}{2\psi_j}\bm r_j^\top\bm r_j,
\]
and
\[
\log m_1(\bm r_j)
=
-\frac n2\log(2\pi)
-\frac12
\left[
n\log\psi_j
+
K\log\left(1+\frac{g}{\psi_j}\right)
\right]
-\frac12
\left[
\frac{\bm r_j^\top\bm r_j}{\psi_j}
-
\frac{g}{\psi_j(\psi_j+g)}
\bm r_j^\top\bm P_X\bm r_j
\right].
\]

\subsection{Full Conditional for $\beta_j$}

When $\gamma_j=0$, $\bm\beta_j=\bm 0_K$. When $\gamma_j=1$, we use
\[
\bm r_j
=
\bm Y_j-\alpha_j\bm 1_n-\bm F\bm\lambda_j.
\]
Then
\[
\bm r_j\mid \bm\beta_j,-
\sim
\mathcal N_n(\bm X\bm\beta_j,\psi_j\bm I_n),
\]
and
\[
\bm\beta_j\mid g,\gamma_j=1
\sim
\mathcal N_K\left\{\bm 0_K,g(\bm X^\top\bm X)^{-1}\right\}.
\]
Thus,
\[
p(\bm\beta_j\mid -)
\propto
\exp\left[
-\frac{1}{2\psi_j}
(\bm r_j-\bm X\bm\beta_j)^\top
(\bm r_j-\bm X\bm\beta_j)
-
\frac{1}{2g}
\bm\beta_j^\top
\bm X^\top\bm X
\bm\beta_j
\right].
\]
Collecting terms gives
\[
\bm\beta_j\mid -
\sim
\mathcal N_K(\bm m_{\beta_j},\bm V_{\beta_j}),
\]
where
\[
\bm V_{\beta_j}
=
\left[
\left(\psi_j^{-1}+g^{-1}\right)
\bm X^\top\bm X
\right]^{-1}
=
\frac{\psi_j g}{\psi_j+g}
(\bm X^\top\bm X)^{-1},
\]
and
\[
\bm m_{\beta_j}
=
\bm V_{\beta_j}
\psi_j^{-1}
\bm X^\top\bm r_j
=
\frac{g}{\psi_j+g}
(\bm X^\top\bm X)^{-1}\bm X^\top\bm r_j.
\]

\subsection{Full Conditional for $\psi_j$}

Let
\[
e_{ij}
=
Y_j(\bm{s}_i)
-
\alpha_j
-
\bm X(\bm{s}_i)^\top\bm\beta_j
-
\bm f_i^\top\bm\lambda_j.
\]
The Gaussian latent regression model gives
\[
\prod_{i=1}^n
p\{Y_j(\bm{s}_i)\mid -\}
\propto
\psi_j^{-n/2}
\exp\left\{
-\frac{1}{2\psi_j}
\sum_{i=1}^n e_{ij}^2
\right\}.
\]
Combining this with $\psi_j\sim\mathcal{IG}(a_\psi,b_\psi)$ yields
\[
\psi_j\mid -
\sim
\mathcal{IG}
\left(
a_\psi+\frac n2,
b_\psi+\frac12
\sum_{i=1}^n e_{ij}^2
\right).
\]

\subsection{Full Conditional for $g$}

Let $q=\sum_{j=1}^p\gamma_j$. Conditional on the active coefficients,
\[
p(\{\bm\beta_j:\gamma_j=1\}\mid g)
\propto
g^{-Kq/2}
\exp\left\{
-\frac{1}{2g}
\sum_{j:\gamma_j=1}
\bm\beta_j^\top
\bm X^\top\bm X
\bm\beta_j
\right\}.
\]
Combining this likelihood contribution with $g\sim\mathcal{IG}(a_g,b_g)$ gives
\[
g\mid -
\sim
\mathcal{IG}
\left(
a_g+\frac{Kq}{2},
b_g+
\frac12
\sum_{j:\gamma_j=1}
\bm\beta_j^\top
\bm X^\top\bm X
\bm\beta_j
\right).
\]

\subsection{Metropolis-Hastings Update for $\phi_j$}

The negative binomial dispersion $\phi_j$ does not have a conjugate full conditional. Its log full conditional, up to terms not depending on $\phi_j$, is
\begin{align*}
\ell(\phi_j)
=
\sum_{i=1}^n
\Big[
\log\Gamma\{C_j(\bm{s}_i)+\phi_j\}
&-
\log\Gamma(\phi_j)
+
\phi_j\log\phi_j \\
&-
\{C_j(\bm{s}_i)+\phi_j\}
\log\{\phi_j+\mu_j(\bm{s}_i)\}
\Big] \\
&+
(a_\phi-1)\log\phi_j
-
b_\phi\phi_j.
\end{align*}
We use a log-scale random-walk proposal,
\[
\log\phi_j^\star
=
\log\phi_j+\varepsilon,
\qquad
\varepsilon\sim\mathcal N(0,\sigma_\phi^2).
\]
Equivalently, $\phi_j^\star=\phi_j\exp(\varepsilon)$. The Metropolis-Hastings acceptance probability is
\[
a(\phi_j,\phi_j^\star)
=
\min\left\{
1,
\exp\left[
\ell(\phi_j^\star)-\ell(\phi_j)
+
\log\phi_j^\star-\log\phi_j
\right]
\right\},
\]
where the final two terms account for the Jacobian of the log-scale proposal.

\section{A Simplified Version for Computational Efficiency}
\label{sub:simplified-versions}






Gene expression datasets including SRT data often contain genes with only a few non-zero counts. The presence of these genes leads to stability issues for the Gibbs sampling. Additionally these are highly unlikely to be spatially varying. We used a simple screening method to eliminate such genes based on their normalized expression values. We rewrite the gene detection problem as a model selection problem as below:
%
\begin{align*}
&\bm{Y}_{j} = (Y_j(\bm{s}_1),Y_j(\bm{s}_2),\dots, Y_j(\bm{s}_n))^{\top} \mid \bm{\beta}_j, \sigma^2 \sim \mathcal{N}_{n}\left(\bm{X}\bm{\beta}_{j}, \sigma^2 \bm{I}_{n} \right)\\
&M_1:\bm{\beta}_{j} \sim \mathcal{N}_p\left(0, g\sigma^2(\bm{X}^T\bm{X})^{-1}\right)\\ &M_0: \bm{\beta}_{j} = \bm{0}_p\\
&\sigma^2 \sim \mathcal{IG}(a,b)
\end{align*}
The advantage of this screening method is that the Bayes factor for model selection has a closed form making the screening process very fast,
\begin{equation}
\label{preselection-model-BF}
BF(M_1,M_0) = \frac{(1+g)^{(n + K - 1)/2}}{(1+g(1-R^2))^{(n-1)/2}}\\
\end{equation}
where $R^2 = 1 - \frac{(\bm{Y}_j - \bm{X}\bm{\beta}_j)^{\top}(\bm{Y}_j - \bm{X}\bm{\beta}_j)}{\bm{Y}_{j}^{\top}\bm{Y}_{j}}$ is the coefficient of determination.  We use the plug-in Empirical Bayes estimate of $g$, i.e.,
$$\widehat{g}_{EB} = \max\left\{\frac{(n - K - 1)R^2}{K(1-R^2)} - 1, 0\right\}.$$
We calculate the Bayes factor for each gene and then eliminate the genes with small values of the Bayes factor.
%
%
%
%
%
%
%
%
%
%
\newpage
\section{Details on Simulation Experiments}
\label{supp:sim-exps}

\begin{table}[H]
\caption{Two simulation settings used to compare the performance of JASPER to competing models SPARK and SpatialDE in the presence of correlated gene expressions.}
\label{tab:simulation-settings}
\centering
\begin{tabular}{|p{\textwidth}|}
\hline
\multicolumn{1}{|c|}{\textbf{Shared Simulation Framework}} \\
\hline
\[
C_j(\bm{s}_i) \sim \text{NB}(1000, p_j(\bm{s}_i)) \quad \text{for } j = 1, \dots, p \text{ and } i = 1, 2, \dots, n
\]
\[
Y_j(\bm{s}_i) = \log \left( \frac{p_j(\bm{s}_i)}{1 - p_j(\bm{s}_i)} \right) = \mu_j(\bm{s}_i) + \epsilon_j(\bm{s}_i)
\]
\[
\bm{\epsilon}(\bm{s}_i) \sim \mathcal{N}(\bm{0}_p, \bm{\Sigma}_{\rho}), \quad \bm{\Sigma}_{\rho} = \begin{pmatrix}
1 & \rho & \dots & \rho \\
\rho & 1 & \dots & \rho \\
\vdots & \vdots & \ddots & \vdots \\
\rho & \rho & \dots & 1
\end{pmatrix}
\]
\\
\hline

\begin{tabular}{p{0.45\textwidth} |p{0.45\textwidth}}
\centering
\textbf{Simulation Setting 1}
\[
\mu_j(\bm{s}_i) = \alpha_j + \sum_{k=1}^5 \beta_{kj} Z_k(\bm{s}_i)
\]
\[
\alpha_j \sim \mathcal{N}(-10, 1)
\]
\[
\beta_{kj} \sim \mathcal{N}(\mathfrak{A}_j\mu_k, 0.1)
\]
\[
\mathfrak{A}_j = 
\begin{cases} 
0 & \text{for non-spatial} \\
\mathfrak{A} & \text{for spatial}
\end{cases}
\]
\[
0 < \mathfrak{A} \leq 1
\]
\[
(\mu_1, \mu_2, \dots, \mu_5) = (2, -2, 3, 1, 4)
\]
&
\centering
\textbf{Simulation Setting 2}
\[
\mu_j(\bm{s}_i) = -10 + \sum_{i=1}^{10} \beta_{kj} T_k(\bm{s}_i)
\]
\[
T_k(\bm{s}_i) = \begin{cases}
Z_k(\bm{s}_i) & \text{for non-spatial} \\
W_k(\bm{s}_i) & \text{for spatial}
\end{cases}
\]
\[
\beta_{kj} \sim \text{Uniform} \{\pm 1, \pm 2, \pm 3, \pm 4\}
\]
\[
Z_k \sim \mathcal{GP}(0, \text{Mat\'ern}(1, 1.5, 0.05))
\]
\[
W_k \sim \mathcal{GP}(0, \text{Mat\'ern}(1, 1.5, 0.5 \mathfrak{A}))
\]
\[
0.01 < \mathfrak{A} \leq 1
\]
\end{tabular}
\\
\hline
\end{tabular}
\end{table}



%
%
%
%
%
%
%
%
%
%
\newpage
\section{Real Data Results}

\begin{table}[H]
\centering
\caption{Comparison of SVG detection methods for the mouse visual cortex dataset generated using STARmap technology.}
\label{tab:starmap}

\small

\begin{tabular}{llccc}
\toprule
\textbf{Category} & \textbf{Metric} & \textbf{JASPER} & \textbf{SPARK} & \textbf{SpatialDE} \\
\midrule

\multirow{4}{*}{SVGs}
& SVGs detected           & 53 & 45 & 22 \\
& Harmonizome overlap     & 19 & 12 & 1  \\
& Original study overlap  & 13 & 11 & 3  \\
& Average Moran's $I$     & 0.73 & 0.61 & 0.41 \\

\midrule

\multirow{2}{*}{Unique SVGs}
& Unique SVGs             & 12 & 4 & 5 \\
& Harmonizome overlap     & 6 & 1 & 0 \\

\bottomrule
\end{tabular}
\end{table}

\begin{table}[H]
\centering
\caption{Biological significance of selected genes uniquely detected by JASPER.}
\label{genes-starmap}
\label{genes-BC}
\label{genes-MOB}

\footnotesize

\begin{tabular}{
p{0.10\linewidth}
p{0.08\linewidth}
p{0.53\linewidth}
p{0.20\linewidth}
}
\toprule
\textbf{Dataset} &
\textbf{Gene} &
\textbf{Biological significance} &
\textbf{Reference} \\
\midrule

\multirow{3}{*}{STARmap}
& \textit{Gad1}
& Encodes glutamate decarboxylase 67, a key enzyme involved in GABA synthesis; plays an important role in nervous system development and function and has been implicated in optic regeneration after injury.
& \cite{ref-gad1} \\

& \textit{Htr3a}
& Associated with the reactivation of the deprived visual cortex and modulation of cortical circuitry.
& \cite{ref-Htr3a} \\

& \textit{Npas4}
& Regulates excitatory--inhibitory balance within neural circuits and contributes to activity-dependent synaptic plasticity.
& \cite{ref-npas4} \\

\midrule

\multirow{3}{*}{\parbox{0.14\linewidth}{Mouse\\Olfactory Bulb}}
& \textit{Kif2a}
& Regulates spindle assembly and the metaphase I--anaphase I transition in mouse oocytes.
& \cite{ref-kif2a} \\

& \textit{Grb2}
& Involved in hepatocyte growth factor-mediated migration of olfactory interneuron precursors through Met--Grb2 signaling.
& \cite{ref-grb2} \\

& \textit{Dlgap4}
& Associated with synapse organization, neuronal signaling, ventricular surface integrity, and neuronal migration during cortical development.
& \cite{ref-dlgap4} \\

\bottomrule
\end{tabular}
\end{table}

\section{Miscellaneous implementation details}

\subsection{A normalization scheme for Gene Expression data}
\label{suppsection:misc:norm}

A number of standard normalization techniques are prelevant for gene expression data. For JASPER-normalized, we used the following scheme as a default.

\begin{equation}\label{normalization}
    Y_j(\bm{s}_i) = \log \left(1 + m\frac{C_j(\bm{s}_i) + 0.01}{N(\bm{s}_i)}\right)
\end{equation}

where $N(\bm{s}_i) = \sum_{j=1}^{p} C_j(\bm{s}_i)$ is the total count at location $\bm{s}_i$, also known as its sampling depth and $m$ is the median of the sampling depths across all locations. 

\subsection{Cubic B-spline basis functions}
\label{supp:cubic-b-spline}

Cubic B-splines are widely used for constructing smooth approximations or interpolations of real functions defined in two or three dimensions. This spline basis ensures continuity up to the second derivative, providing excellent smoothness suitable for various applications like image processing and physical simulations. Mathematically, cubic B-spline basis functions, denoted as \( N_{i,3}(t) \), where \( i \) is the index of the control point and 3 denotes the degree, are defined recursively using the Cox-de Boor formula:

\begin{itemize}
  \item For \( k = 0 \):
  \[
  N_{i,0}(t) = \begin{cases} 
  1 & \text{if } t_i \leq t < t_{i+1} \\
  0 & \text{otherwise}
  \end{cases}
  \]

  \item For \( k \geq 1 \):
  \[
  N_{i,k}(t) = \frac{t - t_i}{t_{i+k} - t_i} N_{i,k-1}(t) + \frac{t_{i+k+1} - t}{t_{i+k+1} - t_{i+1}} N_{i+1,k-1}(t)
  \]
\end{itemize}

For a real function on 2-dimensional space,\( f(x, y) \), the cubic B-spline approximation is given by the tensor product of cubic B-spline basis functions:

\[
S(x, y) = \sum_{i=0}^{n} \sum_{j=0}^{m} c_{ij} N_{i,3}(x) N_{j,3}(y)
\]

where \( c_{ij} \) are the coefficients which can be determined by fitting the spline to known data points or values of the function. Similarly, for a real function on 3-dimensional space, \( f(x, y, z) \), the cubic B-spline representation extends to three dimensions as follows:

\[
S(x, y, z) = \sum_{i=0}^{n} \sum_{j=0}^{m} \sum_{k=0}^{p} c_{ijk} N_{i,3}(x) N_{j,3}(y) N_{k,3}(z)
\]

where \( c_{ijk} \) are the spline coefficients and \( N_{i,3}(x) \), \( N_{j,3}(y) \), and \( N_{k,3}(z) \) are the cubic B-spline basis functions corresponding to each dimension. The control points can be specified explicitly, or chosen to be equally spaced along the spatial coordinates.

\end{document}